# Teaching Physics in the Woods

Frédéric Bouquet, Julien Bobroff, Lou-Andreas Etienne, Clara Vardon

Université Paris-Saclay, CNRS, Laboratoire de Physique des Solides, 91405, Orsay, France.

**Abstract**

We developed a two-day physics class that uses a nearby forest as a teaching location. Using low-cost material, students design and carry out physics projects outside of the usual controlled environment that is a classroom. In this way they come to realize that physics can be used to understand the real world. They organize and present their results in an original format, an exhibit they collectively build. This project is an introduction to the role physics can play in exploring environmental issues, incorporating a sensitive and positive aspect which is important in this time of environmental crisis.

## INTRODUCTION

Universities cannot ignore the environmental crisis and need to adapt their curricula.[1,2] In France, it will be mandatory after 2027 to offer classes covering the notions of sustainability and environmental crisis in every undergraduate program.[3] Physics, as a discipline, offers several relevant approaches to addressing this issue.[1,4,5] At the undergraduate level, one of the most obvious entry points is the concept of energy, as it is already taught at this level and can be used to address various aspects of the environment such as renewable energy or the greenhouse effect.[6]

In May 2022, the international workshop "Reimagine Physics Teaching" brought together university physics teachers to reflect on original ways and the possible evolution of teaching physics.[7,8] One of the outcomes of this workshop was the idea for a new exercise, taking place largely in a forest and letting students look at the forest with the physicist's eye. The goal is to demonstrate the usefulness of physics to tackle larger problems through an engaging and original activity. Building upon this idea we developed a new exercise: in two days, students design and carry out physics experiments from scratch in the forest, analyze their data, and present their results in an exhibit they build together.

# DESCRIPTION OF THE EXERCISE

The objective of this project-based exercise is to let students explore how physics can describe the real world, outside of the laboratory and controlled experiments. We want the students to see how physics can be used to understand nature, even though nature is more complicated than simplified models. This is a first step in understanding what role physics could play in addressing the environmental crisis. We want students to observe the forest, identify a scientific question based on some of these observations, design and carry out an experiment to investigate this question, analyze the data, communicate the results to peer students, and reflect on their place in the environment.

Additional objectives are to develop students' soft skills: group work, sharing and transmission of knowledge, and creativity. Last but not least, we want students to take pleasure in doing physics even though the environmental crisis is a daunting subject.

This exercise was given to 16 undergraduate students from our university who were following a dual physics and geology curriculum. The students were a mix of second year (12) and third year (4) students. It took place in early September, on two consecutive days. A striking point of this exercise is that it partly takes place outdoors, within a forest a few minutes away from the classroom building. A large picnic table next to a trail served as a meeting point for the outdoor activities. An illustrated summary of these two days is provided in the Supplementary Materials.[9]

## Day one: "understanding the forest"

The starting activity is an ice-breaker to let the participants know each other and to set the tone for this exercise. The students are tasked to wander in the forest individually and in silence and to observe the forest. After 15 minutes, they have to bring a keepsake of their wandering to the picnic table and present it to the rest of the class. The teachers also participate in this activity. Then, still around the picnic table, the teachers present the objectives and the organization of the exercise to the students.

Next, the students, in pairs, have to look for ideas for physics projects. To guide their thoughts, they pick three random constraints among a list, for example: sounds / tiny / very large / count / fast / slow / colors / heavy / light / tree / alive / liquid…[9] They have to imagine projects in relation to each of them; other ideas are also welcomed. This phase of exploration is held in the forest, but after sufficient time, everyone gathers in a classroom to make a catalog of all the project ideas. The teachers help to synthesize ideas and to check the feasibility of projects.

Once the catalog is done, each pair of students chooses a project, not necessarily one of their own. They have then the rest of the day to complete it: devise the experimental protocol, perform the measurements in the forest, and analyze their data. To do so, they only have a limited time (about 5 hours), and low-tech equipment such as their smartphone, ropes, rulers, and measuring tapes. Equipment is added for some specific projects (for example a portable Bluetooth speaker for an acoustic study). The students carry out their measurements autonomously in the forest and can return to the building to analyze their data. To close the day, each group presents their results to the rest of the class informally.

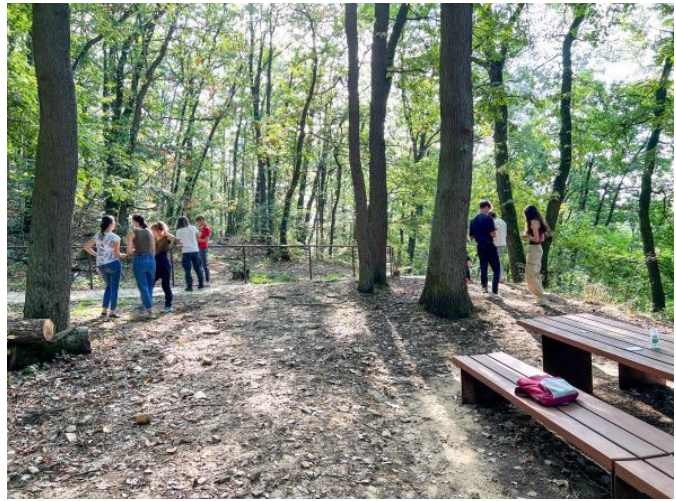

Figure 1: the students are looking for project ideas in the forest.

**Day two: "explaining the forest"**

The goal of the second day is to present all the project results in a shared original exhibit, for which the students are responsible. They are strongly encouraged to look further than a series of posters, and to think of different ways to interact with a public such as mock ups, short activities, small discussions, or quizzes. They are also responsible for the overall organization and design of the space, as well as the different materials that need to be printed (labels, titles, posters, etc.). Their constraint is that the exhibit should be coherent and not simply a juxtaposition of different stands. The start of the day gives the students tools to achieve all of this. They are first invited to explore an "inspiration room", a room that contains references on science, but also design, art, and do-it-yourself projects.[10] They are also given an introduction to fast prototyping, in order to give them a working method for the very short time at their disposal.[9] Rapid prototyping is a method that was first used in software development, but which was later applied to different fields.[11,12] The key idea of this method is to make prototypes very early in the course of a project, even if these prototypes are imperfect and roughly made. The goal is not to have a perfect model that represents the end product, but to obtain a working object with which we can interact and test our ideas.[8] Here we gave the students thirty minutes to build different mock-ups of their ideas for the exhibit using frugal materials such as toothpicks, playmobil sets, stationery, play dough, wooden skewers, cardboard, etc. Students then had three hours to design and build the exhibit, identifying the different tasks that need to be done and dividing them among different task forces. The opening of the exhibit, midafternoon, fixes a strong deadline. Anyone in the building is invited to visit the exhibit and discover the students' work.

Working on an exhibit is unusual for our students; we chose this format to force the students to think outside of the usual lab reports and work on their soft skills. It is a way to highlight their work, and the interaction with the public is rewarding for them, even though the exhibit is short-lived.

The exercise ends with the exhibit opened to other students in the university, followed by a debriefing session, allowing students and teachers to discuss these two days.

**PHYSICS IN THE FOREST**

Different kinds of projects can be carried out in a forest to explore the role of physics:

- measuring a property (e.g. measure the mechanical response of a spider's web);
- estimating a quantity (e.g. estimate the number of trees);
- measuring a distribution (e.g. the distribution of the height of trees);
- testing a law of physics (e.g. the fall of bodies);
- testing a scaling law (e.g. diameter of trees according to their height);
- building an experiment (e.g. build a frugal microscope).

This diversity gives a wide range of possible projects. Even though we want students to define their own projects, we checked the feasibility of this method of teaching physics by playing the role of students beforehand: we made a list of possible projects and tested out some of them. For example, we measured the diameter of a spider silk using a red laser pointer and observing the interference pattern of the illuminated silk. We found an interference pattern consistent with a thread of a 9 μm diameter, with a 10% uncertainty. This is in the expected 5–10 μm diameter range;[13] we also measured a dusty indoor cobweb and found a larger value (22-μm diameter) presumably due to the dust thickening the thread. Manipulating spider silk is a bit delicate, but not that difficult. A student project could be to study whether the diameter of the silks is constant everywhere on a given web, or whether some threads are wider (and possibly stronger) than others. Another project could be to compare outdoor and indoor spiderwebs to investigate if the diameters of the threads are different.

A second example is the measure of the Young's modulus of tree branches. Young's modulus is a parameter that will limit the maximum height a tree can grow before buckling appears.[14] We measured the Young's modulus of a branch (70 cm long, with a 1-cm diameter) by using different techniques: either by applying a force at the extremity and measuring the deflection of the branch, or by measuring its natural oscillation frequency (see the Supplementary Materials[9] for the details of the equations and hypothesis). When we attached a 0.45 kg water bottle at the extremity of the branch, a 5-cm deflection was measured, corresponding to a 2 GPa Young's modulus, with a 20% uncertainty due to the deflection measurement. The natural measurement frequency can be measured by different methods; video analysis, detection of the mechanical vibrations with a smartphone accelerometer in contact with the base of the branch, or with a smartphone magnetometer if a small magnet is attached to the branch. All three methods led to a natural frequency of 11 Hz with a precision of 5%. This is consistent with a Young's modulus of 7 to 8 GPa if a wood density of 500 kg/m$^3$ is assumed. Even though the same branch was used for these measurements, the results are significantly different. The fact that the cantilever models do not exactly describe the system can be caused by various factors; for example, the branch was modeled as a perfect 1-cm diameter cylinder, which it was not. A student project could be to check the scaling laws predicted by the models when varying the load, or the length and diameter of the branch.

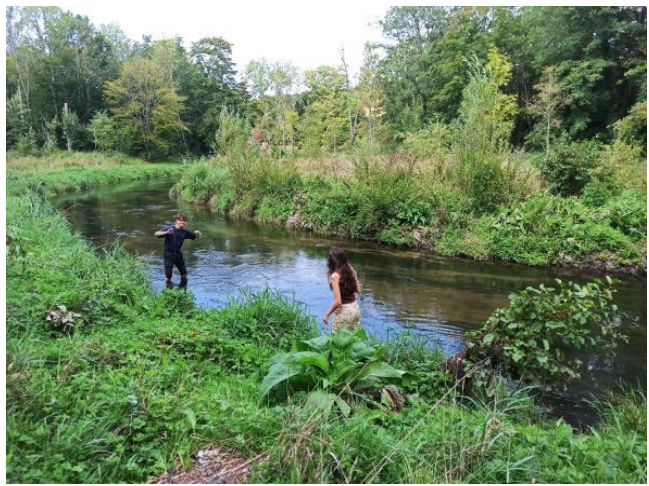

Figure 2: a group of students is studying the flow of a river.

We made a list of experiments to be made available to students lacking ideas for projects which is supplied as Support Material for this article.[9] During the actual exercise, we did not need this list, and the students came up with their own ideas, among which they picked:

- the study of the fall of acorns (Does the free fall model apply to acorns? Can the fall of many acorns kill a passerby?);
- the study of the flow of a river (Compare different ways of measuring or estimating the flow of a river. Is the flow homogeneous across a river cross-section?);
- the optical magnification of a drop of water on a leaf (How does the size of the drop, considered as a lens, affect the magnification.);

- the relationship between the size of the trees and the slope of the ground on which the trees grow (Is there a correlation between these two parameters?);
- measuring the flexibility of a tree or a branch (Measure an averaged Young's modulus modeling tree branches as cantilevers.);
- sound attenuation in the forest (How does the tree surface density affect the sound attenuation?).
- relation between the tree surface density and tree height (Is there a correlation between these two parameters?).

Compared to students' labs in controlled environments, prepared and verified beforehand by experienced teachers, these projects often do not provide clear results. This is deliberate on our part. It helps students understand that a classroom experiment is a carefully contrived situation. Furthermore, they realize that even with the limited time and material at their disposition, they can obtain relevant information; the fact that more advanced experiments would be needed to go further is then discussed.

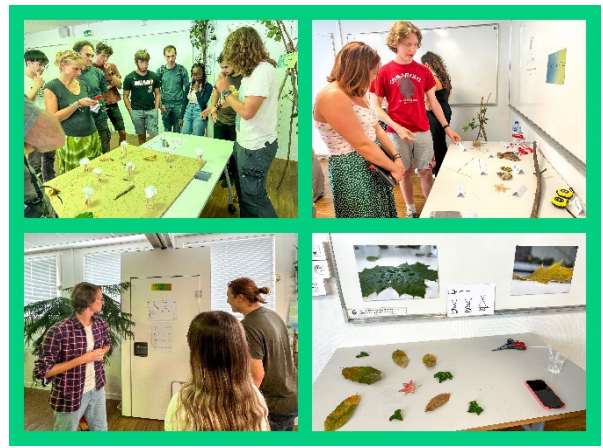

Figure 3: the students present their results in a public exhibit they collectively build.

Letting students develop their own project is an effective way to engage them; they are conducting their own study, not their teacher's. One consequence of this freedom is that some of the projects chosen by the students may have only a remote connection to the notions of environment, sustainability and climate. It should be possible to guide the students a bit more in that direction by letting the students think about how these notions affect the forest and could be linked to their projects. The project conducted by a pair of students during our exercise, looking for a correlation between tree size and the slope of the ground, is a good example of a project that moves away from pure physics into environmental physics.

**STUDENTS' FEEDBACK**

A brief online survey was sent to collect students' feedback, in addition to the oral debriefing. It contained a few questions, some quantitative, and some open ended. The small number of students (16) does not allow for a solid statistical analysis, but a qualitative analysis can be performed.

As often during a novel teaching method,[15] the general feedback from the students is positive. In answer to the survey question "What did you think of this exercise?" the average of the student's answers was 6.1 on a scale from 1 ("I did not enjoy it") to 7 ("I loved it"), and the statements collected during the debriefing or open-ended questions are positive. For example: "*another point of view on equations and physics*" "*Feeling good*" "*Satisfaction at having been able to perform a scientific project from A to Z*" "*we are completely immersed in a social and scientific experiment and it feels good!*" "*I really loved it*".

Regarding physics learning, to the question "Do you feel you have learned more or less than during the same number of hours of classical lab class?", the average answer was 4.6 on a scale from 1 ("much less") to 7 ("much more"). In other words, the students believe they learned about as much during this exercise as during a classic lab experience.

Several main themes can be found in the debriefing and in the students' open-ended responses:

- teamwork ("*getting organized as a team*" "*This lesson gave me knowledge about […] teamwork*" "*teamwork can be very effective if you get organized*" "*teamwork is not easy because we all have different ideas and we don't agree so easily*");
- freedom and autonomy ("*The freedom we had*" "*choose the theme we want to work on ourselves, and set up our own experiences.*")
- the link between theory and reality ("*Science is everywhere, you just have to make a little effort to notice it.*" "*another point of view on equations and physics*" "*a more concrete aspect of research and how we can apply our knowledge*" "*Science is not just long and incomprehensible calculations*")
- the time constraint ("*The time constraints. […] it was an adrenaline rush. It forced us to work fast (in a good way)*" "*stress to finish […] the exhibit*")
- the low-cost aspect of the tools ("*we can accomplish much more than we think with little equipment*" "*there is no need for very sophisticated equipment to do science*" "*with little we can do a satisfying job*" "*one can experiment from very little*")
- the pleasure of being in the forest ("*Being outside and looking at nature through physics*" "*Telling the forest*" "*More forest*")

Having themes in the students' feedback such as teamwork, autonomy, and the link between theory and reality is not surprising since project-based learning is known to have a strong effect on these subjects.[16–18] The low-cost aspect of the material also came up quite often, in a positive way. The students are not put off by this low-cost approach, compared to the usual lab instruments they use in other lab classes. This approach gives the students more freedom than they would have with more precise, but more fragile and more expensive instruments, and is justified by the fact that the focus of this exercise is on the process (how to do physics in real conditions) more than on the precision of the results. This is an approach that we had already successfully tested in another context.[18]

Usually, university instruction taking place in a forest either focuses on the forest (such as management or exploitation of the forest), or uses it for field work in direct relationship to the topic of the exercise (field trip for classes of biology, ecology, or biodiversity).

Our approach is different as we use the forest as a place to teach general physics. It is actually not that different from the Forest Schools.[19,20] Forest School is a form of education particularly developed in Scandinavia and the UK, mainly for kindergartens and elementary schools. It's an outdoor education, mostly woodland, that focuses on learning by doing and hands-on activities. Albeit for the obvious difference in age group, we share the idea of a separate learning space which offers more freedom, being removed from the physical constraints of the classroom.

Another theme appears in the open answers of the students, with a negative aspect: the presence of mosquitoes during the two days of teaching. Despite being somewhat anecdotal, these answers underline an important point: unlike a classroom, where all the technical aspects can be controlled, going outside introduces some unpredictability. The weather is an obvious example of an uncontrollable factor that can impact an outdoor exercise, and it is prudent to plan ahead for different teaching scenarios depending on the conditions.

## ADAPTING THE FORMAT

Is this exercise transposable to other universities, given adaptations to the local context? The salient points seem to us to be the following:
- low-cost experiments outside the labs;

- let students figure out an interesting problem to investigate;

- the autonomy and freedom they have to take charge of their project;

- the creative presentation of the results.

The forest is not that important as such. It serves as a natural setting that students explore, but

Nature is not limited to the forest, and this exercise could easily be adapted to other places: a meadow, an urban park, or the streets of a city. For example, we have also created a list of experiment ideas that could be conducted in an urban setting or during a subway ride (see supplementary materials).

The two consecutive-day format can also be adapted. We appreciate this format for the dynamic it creates with the students who are focused on their project without other academic concerns. It is easy to consider different formats: a short one-day version, or conversely a longer version, with more elaborate projects. One could also conceive hybrid solutions mixing more classical exercise on environmental physics and sessions in the forest giving students the opportunity to apply knowledge they would just have learned.

## CONCLUSION

We have designed an exercise that presents physics as a useful tool to understand and investigate the real world. While the question of the environmental crisis is clearly in the background of this exercise, the setting allows students to take pleasure in doing science and not address the climate crisis directly. This exercise is project-based; its originality is that part of it takes place outside, in the forest near the classroom buildings. It is complementary to classes with a content directly related to climate change, but it brings a more sensitive and positive dimension. We are also in the process of designing a popularization video to stage this approach to physics in the forest.

This format could be adapted to other experimental disciplines, particularly with life sciences. In particular we plan to introduce an interdisciplinary component to our own teaching; physics is not the only tool for understanding Nature, and using a multidisciplinary approach is preferable when studying environmental problems.[5] The same exercise with supervisors coming from different disciplines (chemistry, biology, ecology) would offer more diverse and richer views to the students and would naturally open the projects towards the subjects of environment, sustainability and climate.

## ACKNOWLEDGMENT

We would like to thank Fabienne Bernard, Hervé Caps, Marina Carpineti, Joël Chevrier, Francesca Chiodi, Ulysse Delabre, Fun-Man Fung, Jean-Michel Geneveaux, Natacha Krins, Denis Terwagne, Benjamin Vest, and Maëlle Vilbert who participated in the group that reflected on this teaching during the international workshop "Reimagine Physics Teaching" in May 2022. We also thank the Research Action Chair on Educational Innovation from Paris-Saclay, held by Martin Riopel and financed by the École Universitaire de premier cycle de l'université Paris Saclay. This chair exists thanks to Jeanne Parmentier, Isabelle Demachy, and Martin Riopel, and benefits from the monitoring and methodological support of Patrice Potvin, Christian Bégin, Diane Leduc, Geneviève Allaire-Duquette, and Marine Moyon. This work has been partially supported by the Chair "Physics Reimagined" led by Paris-Saclay University and sponsored by AIR LIQUIDE and Crédit Agricole – CIB, and by the Institut Villebon Georges Charpak.

**BIBLIOGRAPHY**

[1]E. Boeker., R. van Grondelle, and P. Blankert, "Environmental physics as a teaching concept," European Journal of Physics, 24(5), S59–S68 (2003).

[2]K. Forinash, "A few ideas for teaching environmental physics," Physics Education, 51(6), 065024 1–12, (2016).

[3]J. Jouzel, "Sensibiliser et former aux enjeux de la transition écologique et du développement durable dans l'enseignement supérieur," working group conclusion submitted on 02/16/2022 to the French Minister of Higher Education and Research, (2022). Available at https://www.enseignementsup-recherche.gouv.fr/fr/remise-du-rapport-sensibiliser-et-former-aux-enjeux-de-la-transition-ecologique-et-du-developpement-83903

[4]H. C. Busch, "Using Environmental Science as a Motivational Tool to Teach Physics to Non-science Majors," The Physics Teacher, 48(9), 578–581, (2010).

[5]D. A. Kimori and G. Roehrig, "Environmental topics in physics by inquiry course: Integration models used by physics teachers," The Qualitative Report, 26(5), 1601–1617, (2021).

[6]K. Forinash and B. Whitten, "Resource letter TEP-1: resources for teaching environmental physics," American Journal of Physics, 87(6), 421–432, (2019).

[7]J. Bobroff, "The 'anti-conference', a collaborative way to create innovative teaching methods," The Conversation. https://theconversation.com/the-anti-conference-a-collaborative-way-to-create-innovative-teaching-methods-190504 , (2022, September 14).

[8]B. Vest, *et al.*, "Reimagine Physics Teaching: a workshop designed to sparkle exchanges and creativity," les cahiers de l'Institut Pascal, accepted for publication.

[9]Supplementary Material is found at [url to be inserted by AIPP].

[10]As an example, the inspiration room is presented here: https://hebergement.universite-paris-

saclay.fr/supraconductivite/references-sur-la-foret/?lang=en

[11]T. S. Jones, and C. R. Richey, "Rapid prototyping methodology in action: A developmental study," Educational Technology Research and Development, 48, 63– 80, (2000).

[12]E. K. Nixon, and D. Lee, "Rapid prototyping in the instructional design process," Performance Improvement Quarterly, 14(3), 95-116, (2001).

[13]T. A. Blackledge, R. A. Cardullo, and C. Y. Hayashi, "Polarized light microscopy, variability in spider silk diameters, and the mechanical characterization of spider silk," Invertebrate Biology, 124(2), 165–173, (2005).

[14]T. McMahon, "Size and shape in biology," Science, 179(4079), 1201-1204, (1973).

[15]J. Hattie, *Visible learning for teachers: Maximizing impact on learning*, Routledge, (2012).

[16]P. C. Blumenfeld *et al.*, "Motivating project-based learning: Sustaining the doing, supporting the learning," Educ. Psychol. 26(3–4), 369–398 (1991).

[17]C. Reverdy, "Des projets pour mieux apprendre ?," Dossier d'actual. Veille Anal. 82, 1–24 (2013); available at https://hal.archives-ouvertes.fr/hal-01657236/document

[18]F. Bouquet, J. Bobroff, M. Fuchs-Gallezot, and L. Maurines, "Project-based physics labs using low-cost open-source hardware," American Journal of Physics, 85(3), 216–222, (2017).

[19]F. Harris, "Outdoor learning spaces: The case of forest school," Area, 50(2), 222-231, (2018). https://doi.org/10.1111/area.12360

[20]F. Harris, "Forest school," CABI Reviews, (2022). https://doi.org/10.1079/cabireviews202217041